\documentclass[a4paper,11pt]{article}
\pdfoutput=1

\usepackage{jheppub,fancyhdr,graphicx,epstopdf,color,amsmath,cases,slashed,hyperref,subfigure}
\usepackage{comment}

\newcommand{\beq}{\begin{eqnarray}}
\newcommand{\eeq}{\end{eqnarray}}

\def\ltap{\ \raise.3ex\hbox{$<$\kern-.75em\lower1ex\hbox{$\sim$}}\ }
\def\gtap{\ \raise.3ex\hbox{$>$\kern-.75em\lower1ex\hbox{$\sim$}}\ }

\def\be{\begin{equation}}
\def\ee{\end{equation}}
\def\bea{\begin{eqnarray}}
\def\eea{\end{eqnarray}}

\newcommand{\gluino}{\tilde{g}}

\newcommand{\squark}{\tilde{q}}

\definecolor{red1}{cmyk}{0,1,1,0.3}

\title{\boldmath A novel signature for long-lived particles at the LHC}

\preprint{LAPTH-021/17, IPPP/18/87}

\author[a,b]{Shankha~Banerjee}
\author[a]{\!\!, Genevi\`{e}ve~B\'{e}langer}
\author[c]{\!\!, Biplob~Bhattacherjee}
\author[a]{\!\!, Fawzi~Boudjema}
\author[c]{\!\!, Rohini~M.~Godbole}
\author[d]{\!\!\!, Swagata~Mukherjee}

\affiliation[a]{LAPTh, Universit\'e Savoie Mont Blanc, CNRS, BP~110, F-74941 Annecy-le-Vieux, France}
\affiliation[b]{Institute for Particle Physics Phenomenology, Department of Physics, Durham University, Durham DH1 3LE, United Kingdom}
\affiliation[c]{Centre for High Energy Physics, Indian Institute of Science, Bangalore 560012, India}
\affiliation[d]{III. Physikalisches Institut A, RWTH Aachen University, Otto-Blumenthal-Str. 16, 52074 Aachen, Germany}

\emailAdd{shankha.banerjee@durham.ac.uk}
\emailAdd{belanger@lapth.cnrs.fr}
\emailAdd{biplob@iisc.ac.in}
\emailAdd{boudjema@lapth.cnrs.fr}
\emailAdd{rohini@iisc.ac.in}
\emailAdd{mukherjee@physik.rwth-aachen.de}
\abstract
{In contrast to the decay products ensuing from a fast moving particle which are collimated along the original direction of the parent, those from a slow moving particle are distributed over a wide region. In the context of searches for heavy long-lived particles (LLP) at the Large Hadron Collider (LHC), we quantitatively demonstrate, using a few benchmark models, that objects which emerge from a secondary vertex due to the decay of an LLP at the TeV scale can be at large angular separations with respect to the direction of the parent LLP. A fraction of the decay products, the backward moving objects (\textit{BMO}s), can even go in the backward direction. These will give rise to striking signatures in the detectors at the LHC as these particles will traverse  different layers of the detector {\it outside-in} towards the direction of the beam-pipe. Based on a simple geometrical modelling of the detector, we give examples of how this effect translates into the fraction of energy deposited in the tracker, from particles coming as far as from the hadron calorimeter, as well as those that could be entering from outside the detector into the muon chamber. The largest effect is from LLP candidates that come to rest inside the detector, such as the stopped $R$-hadrons. But the results are promising even in the case of not so heavy LLPs and/or when  some of the available energy is carried by a massive invisible daughter. This urges us to look more in details at these unusual signatures, taking into account the particularities of each layer that constitutes the detector. From  the \textit{BMO} perspective, we review how each layer of the detector could be exploited and what improvements can be made to enhance the shower shapes and the timing information, for instance. We also argue that the cosmic ray events, the most important background, can be easily dealt with. 
}

\usepackage{array}
\newcolumntype{+}{>{\global\let\currentrowstyle\relax}}
\newcolumntype{^}{>{\currentrowstyle}}
\newcommand{\rowstyle}[1]{\gdef\currentrowstyle{#1}%
#1\ignorespaces
}

\begin{document}

\date\today

\maketitle
\flushbottom

\section{Introduction}
\label{sec:intro}
Long-lived massive particles (LLP) are predicted in many extensions of the Standard Model (SM) that address the hierarchy problem~\cite{Giudice:1998bp, Arvanitaki:2012ps}, naturalness~\cite{Chacko:2005pe, Burdman:2006tz, Craig:2015pha, Curtin:2015fna, Chacko:2015fbc}, the baryon-antibaryon asymmetry in the universe~\cite{Cui:2014twa}, and dark matter (DM)~\cite{Co:2015pka, Godbole:2015gma, Khoze:2017ixx, Garny:2017rxs}  including feebly interacting particles~\cite{Co:2015pka, Evans:2016zau, Banerjee:2016uyt} and asymmetric DM~\cite{Bai:2013xga}. They can also impact the phenomenology of the Higgs boson~\cite{Maiezza:2015lza,Dev:2017dui}. 

The LLP's long lifetime can be due to: {\it i)} a much reduced phase space resulting from a small mass splitting between the LLP and one of its decay products (as found in AMSB models~\cite{Randall:1998uk}) or {\it ii)} a suppressed coupling that controls the dominant decay, examples include SUSY models with a gravitino DM~\cite{Dimopoulos:1996vz, Asai:2011wy, Jung:2015boa, Allanach:2016pam}, $R$-parity violating (RPV) scenarios~\cite{Graham:2012th,Evans:2016zau}, and models containing a hidden sector that is weakly coupled to the SM via some mediator~\cite{Strassler:2006ri, Han:2007ae}. The suppressed (effective) coupling can also result from the fact that the main decay proceeds through a mediator whose mass is very high compared to the LLP ({\it e.g.} $R$-hadrons as bound states of gluino ($\gluino$) in split SUSY with very heavy squarks~\cite{Kraan:2004tz, Mackeprang:2006gx}).  

The characteristic long life-time, ranging between 100 picoseconds to a few nanoseconds, of these massive particles translates, at the experimental level, to a characteristic feature that has been exploited in many analyses by the ATLAS, CMS and LHCb Collaborations at the Large Hadron Collider (LHC): they decay at some distance (tens to hundreds of centimeters) from the interaction point. When an LLP is produced at the primary vertex and decays at a certain distance inside the detector, \textit{i.e.}, at the secondary vertex, a typical search method is to identify this displaced vertex~\cite{Sirunyan:2017ezt}, as has been pursued for neutral LLPs~\cite{CMS:2014wda, Aaij:2014nma, CMS:2014hka, Khachatryan:2014mea, Aad:2015rba, Aaij:2016xmb}. More specific or tailor-made analyses exploit the location of the secondary vertex within a particular layer of the detector (tracker, electromagnetic calorimeter or ECAL, hadronic calorimeter or HCAL, or muon chamber), the nature of the LLP (charged or neutral)  and  the signatures of the decay products. For instance some charged LLPs are identified by leaving only some visible tracks in the inner layer of the tracker before seeming to disappear in the outer layers as the decay products go undetected because they are either weakly interacting neutral particles and/or too soft. Disappearing tracks~\cite{CMS:2014gxa} and tracks with kinks belong to this category~\cite{Barate:1999gm, Asai:2011wy, Jung:2015boa, Curtin:2018mvb}. Strategies to look for charged particles that are long-lived enough to escape the entire detector~\cite{CMS-PAS-EXO-16-036, Allanach:2002nj} have also been designed. As in the case of some neutral LLPs, the inner tracker may not be of much use and one may rely on the muon spectrometer~\cite{Aad:2014yea}. Especially in the case of fast LLPs giving rise to collimated final states, leptonic decay products, that materialise in the HCAL or the outer edges of the ECAL, may be reconstructed as jets (lepton jets) with a peculiar energy deposition~\cite{Aad:2012kw, Aad:2014yea, Aad:2015asa, ATLAS:2016jza}. For (neutral) LLPs whose decay products consist of photons, exploiting the capabilities of the ECAL within a displaced vertex reveals the LLP through photons that are non-pointing (to the primary vertex) or delayed (compared to prompt photons)~\cite{Aad:2014gfa}. Other scenarios with many final state decay products can rely on a few overlapping displaced vertices (emerging jets~\cite{Schwaller:2015gea}). 

A common underlying feature of most of these LLP searches  is that they are based on {\em inside-out} analyses, looking at the ordered sequence of events going successively from the inner layers (and sublayers) of the detectors to the outer layers, that is from the interaction point (or the beam) to layers in the tracker, to the ECAL, the HCAL and the muon chamber. This is the normal sequence even in the case of  `standard' beyond the standard model (BSM) particle searches. This seems to be the logical sequence, as when a certain heavy particle is produced at the interaction point, it moves forward in time and outward from the beam pipe through the successive inner layers of the detector. What we would like to underline in this paper is that there are instances where an {\em outside-in} approach (at least between two regions of the above ordered sequence), starting from the location of the secondary vertex~\cite{Sirunyan:2017ezt}, is possible and that it should be fully exploited since the signatures are striking with little standard model background. What we will take advantage of is the fact that while the LLP is travelling {\em inside-out}, away from the beam, a proportion of its decay products, those being emitted in the opposite (backward) direction with respect to the direction of the LLP, seem to move inward and hit {\em outside-in} some of the layers or/and sublayers of the detectors. In the latter and in the particular case of jets as decay products, it can also happen that these jets emanating from a  displaced vertex located in the HCAL, are deviated, compared to prompt jets emerging form the production vertex. As a result, they hit multi-towers of the HCAL contrary to the prompt jets that hit only one tower of the HCAL. Such a manifestation is akin to the case of non-pointing photons listed in the previous paragraph. These scenarios are in sharp contrast to the production of particles that experience a large boost and therefore carry all their decay products in their original direction. Since the proportion of daughter particles from the decay of massive LLPs that may experience an {\em outside-in} trajectory is of key importance, section~\ref{sec:sec2} is dedicated to a detailed study of generic scenarios according to production modes, decay signatures and masses, as well as  the possible influence of spin. As expected, the heavier and hence slower the LLP, the larger the proportion of backward daughters should be. At the LHC, the range of masses that can be exploited is also quite wide. Objects, such as stopped hadrons, which represent  particles that lose all their energy and decay after coming to rest within the detector~\cite{Khachatryan:2015jha, Aad:2013gva, Abazov:2007ht}, are an extreme case of slow moving objects and therefore benefit from the general observations we make in this paper.  Although in order to solidly quantify the benefits of our approach requires implementing the details of the detector geometry and the triggers, we nonetheless conduct a simple simulation in section~\ref{sec:sec3}. In section~\ref{sec:sec4}, we discuss the present and future experimental possibilities to deal with such new signatures, in particular the present limitations of the detectors and what future implementations can be added. A summary of the salient points of our paper together with some recommendations are left for the conclusion.

\section{Angular characteristics of the decay products for pair produced heavy particles}
\label{sec:sec2}

Our analysis starts by looking at the angular features of  the final products of the decaying heavy particles, $X$, that are pair produced at the LHC. In particular, we have in  mind the alignment of the daughter particles with respect to the original direction of the parent particle, $X$. Our choice of signatures is based on typical examples of LLP scenarios. However, we will first perform a model-independent investigation in order to find out whether the specific underlying model-dependent dynamics have important roles in the salient features that we want to emphasise. The daughters can be \textit{massless} quarks, $q$, or heavy invisible particles, DM, which may satisfy the properties of dark matter. We consider the following four distinct possibilities.

\begin{itemize}
\item[$\bullet$]  $X \to q \; q$  \\
The decay into a pair of a massless quarks is motivated by, \textit{e.g.}, $R$-Parity Violating (RPV) decays of a squark in supersymmetry, $\squark \to q \; q $, and has connection with $R$-hadrons. Another example is a slepton $\tilde{l}$ decaying into a pair of quarks through RPV, $\tilde{l} \to qq$. We will use the latter for our simulation of this class of scenarios which we henceforth refer to as {\it 2BM0}, corresponding to two-body \textit{massless} final state. One should keep in mind that the production is of the Drell-Yan kind, being initiated by quarks. Having considered scalar mother particles in this example, there is no spin correlation to worry about. 

\item[$\bullet$] $X \to q \; q \; q$ \\
This case is also within the purview of RPV. A prototype which we will use in our simulation is the three-body decay of a neutralino into quarks, $\tilde{\chi}^0_{1} \to q \; q \; q$. This is thus defined as our {\it 3BM0} class. Once again, the production is quark initiated but the effects of spin correlation may not be negligible.

\item[$\bullet$]  $X \to q  \; \textrm{DM}$\\
Here we consider decays of the LLP into a heavy neutral invisible particle, $\textrm{DM}$, which may be a dark matter candidate, alongside a light quark. This scenario arises in $R$-parity conserving SUSY processes, \textit{viz.} $\squark \to q \tilde{\chi}_1^0$, or the radiatively induced $\gluino \to g \tilde{\chi}_1^0$, where $\tilde{\chi}_1^0$ is the lightest neutralino, which can potentially be a DM candidate. Our prototype here is based on the lightest sbottom decay into a bottom quark and a neutralino, $\tilde{b}_1 \to b \tilde{\chi}_1^0$. This class will be termed {\it 2BM}. In our prototype of this class of processes, the production is dominantly gluon induced. 

\item[$\bullet$] $X \to q \; q \; \textrm{DM}$ \\
The final class of processes that we will consider is the {\it 3BM}. This may be represented in $R$-parity conserving SUSY scenarios by the three-body decay $\gluino \to q \bar{q} \tilde{\chi}_1^0$. Our simulation here will be based on this decay. As such, the production process here is also dominantly gluon induced. We will use this example to study the effects of spin correlation, later in this section.
\end{itemize}

\begin{figure}[tbhp]
\subfigure[]
{
 \includegraphics[height=4cm,width=8.0cm]{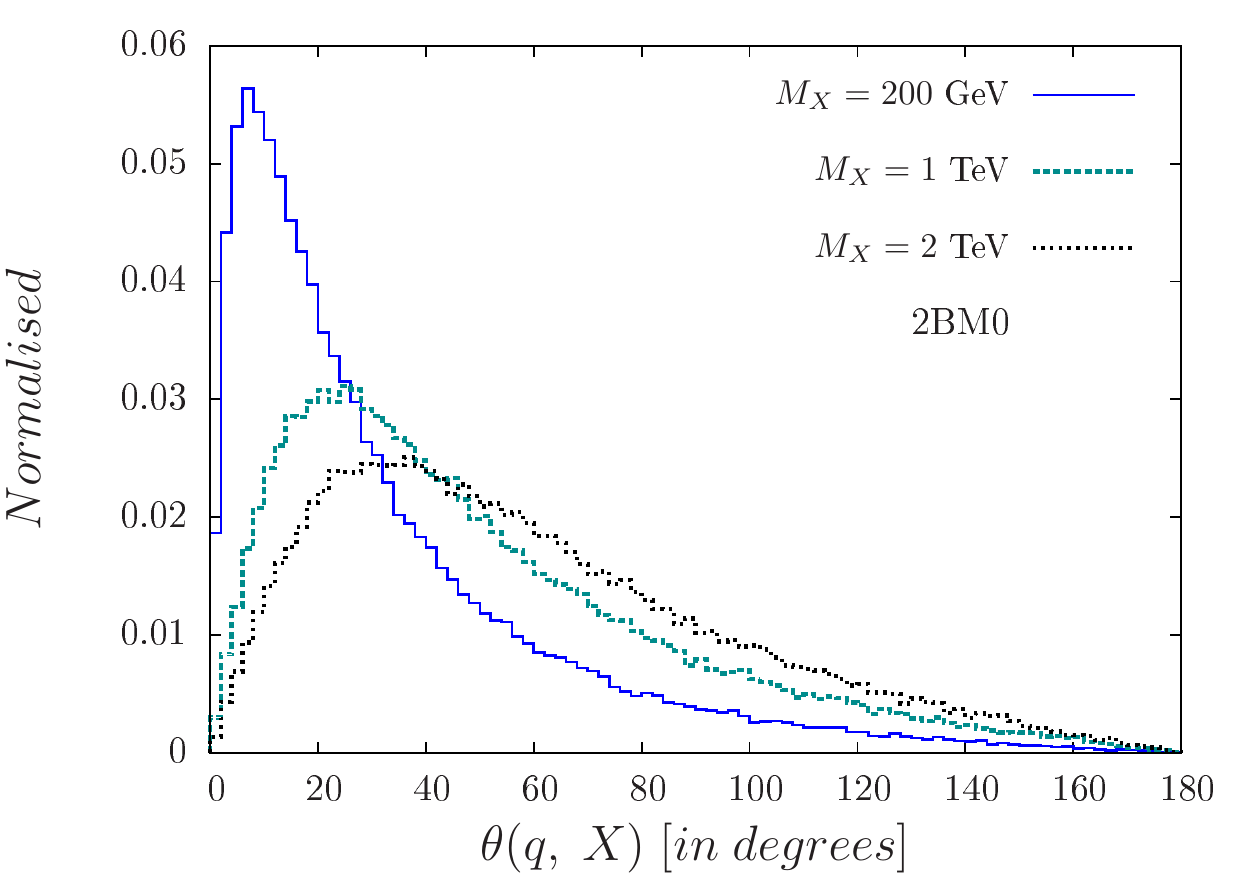}
}
\subfigure[]
{
\includegraphics[height=4cm,width=8.0cm]{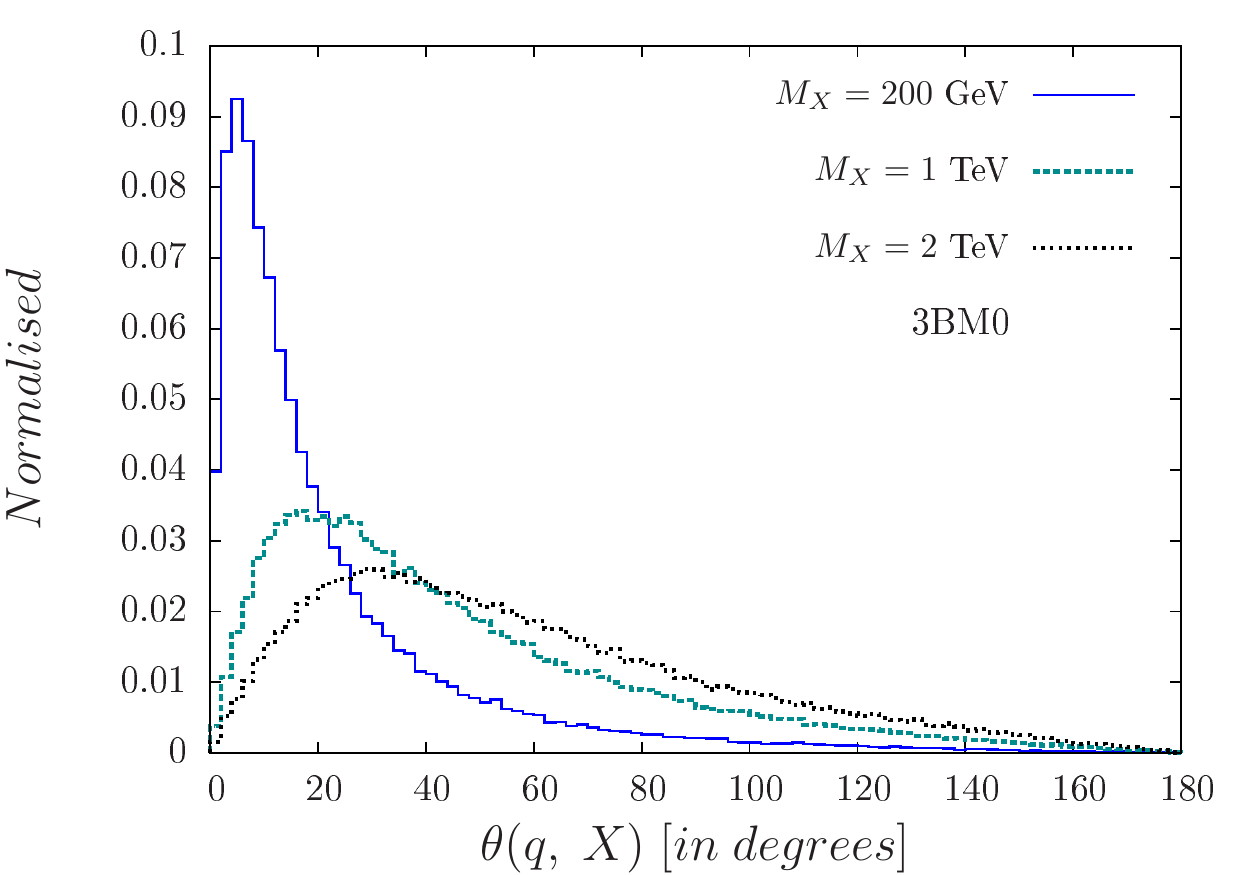}\\
}
\subfigure[]
{
 \includegraphics[height=4cm,width=8.0cm]{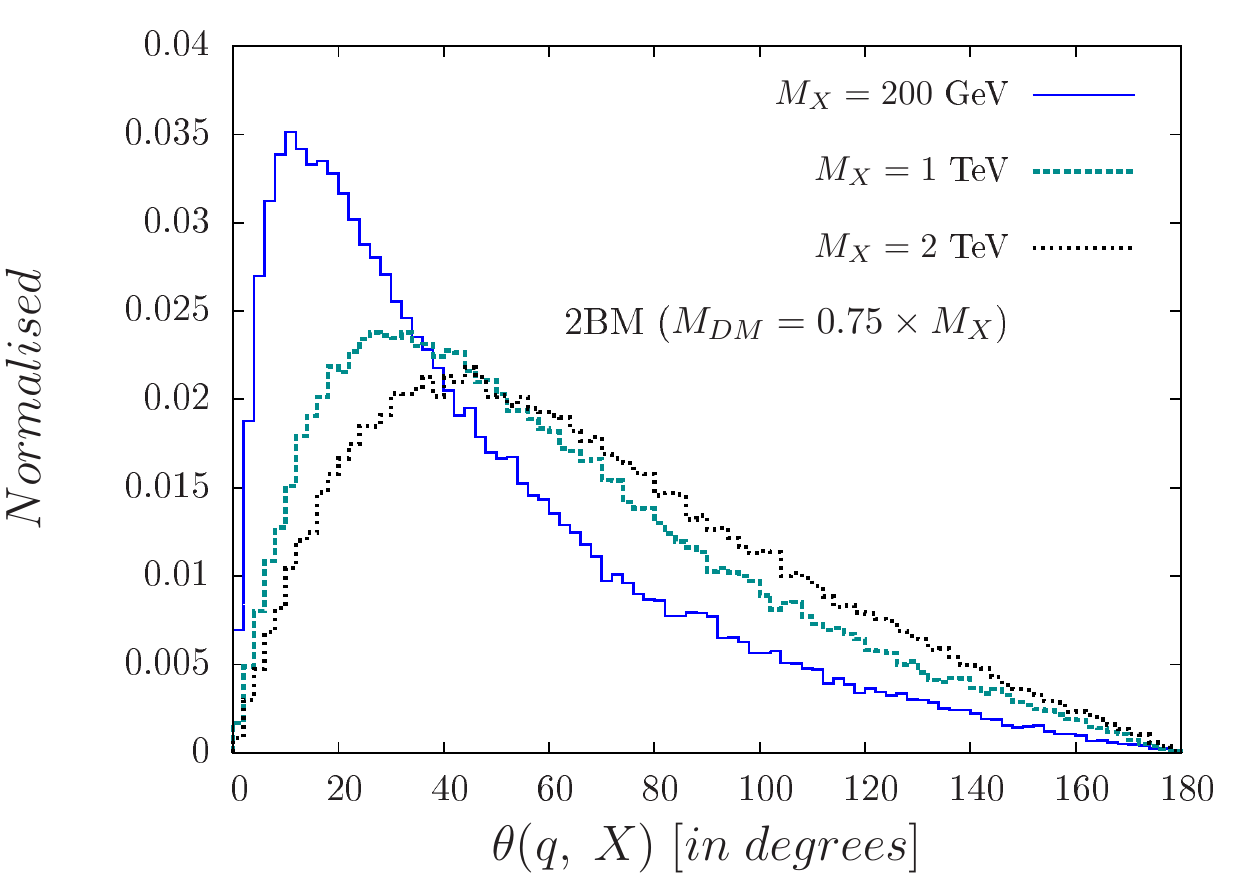}
}
\subfigure[]
{
\includegraphics[height=4cm,width=8.0cm]{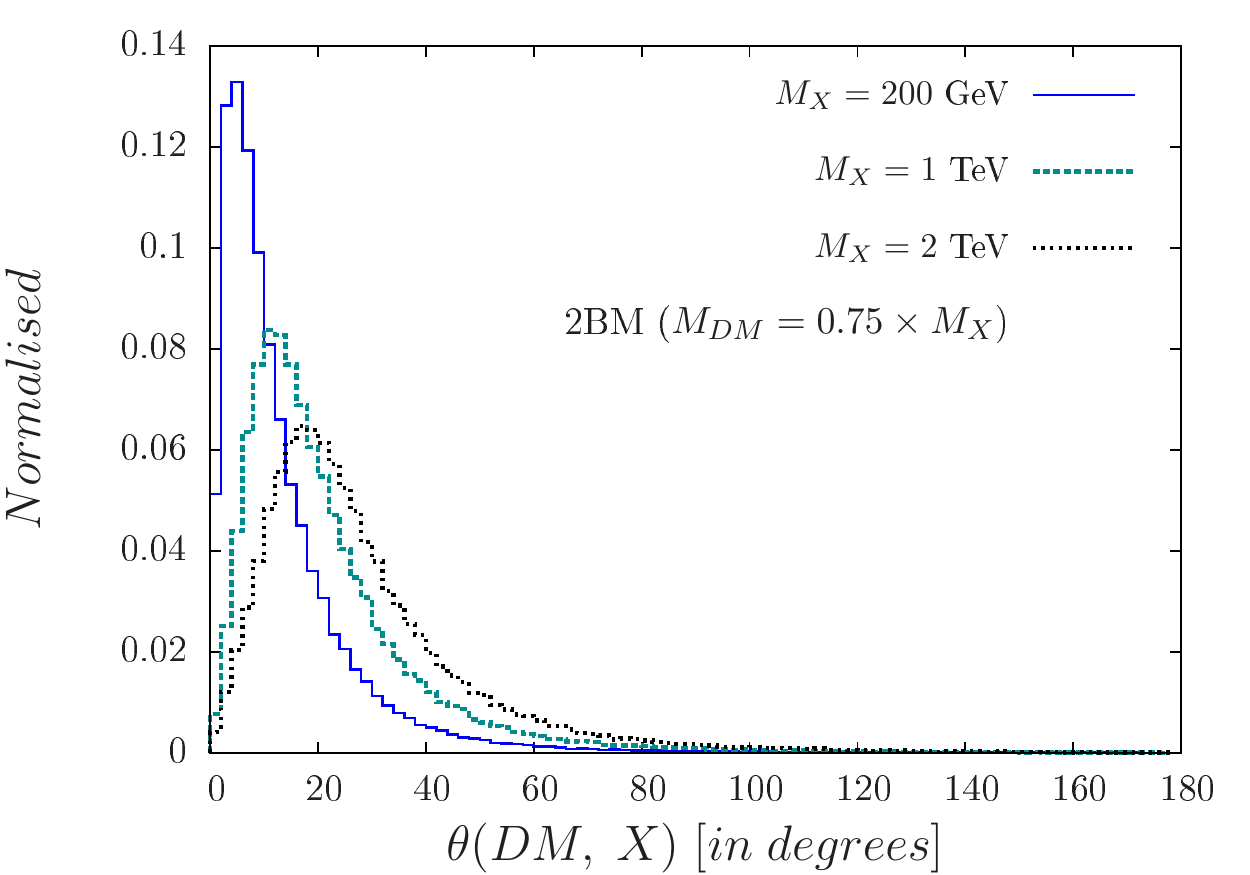}\\
}
\subfigure[]
{
 \includegraphics[height=4cm,width=8.0cm]{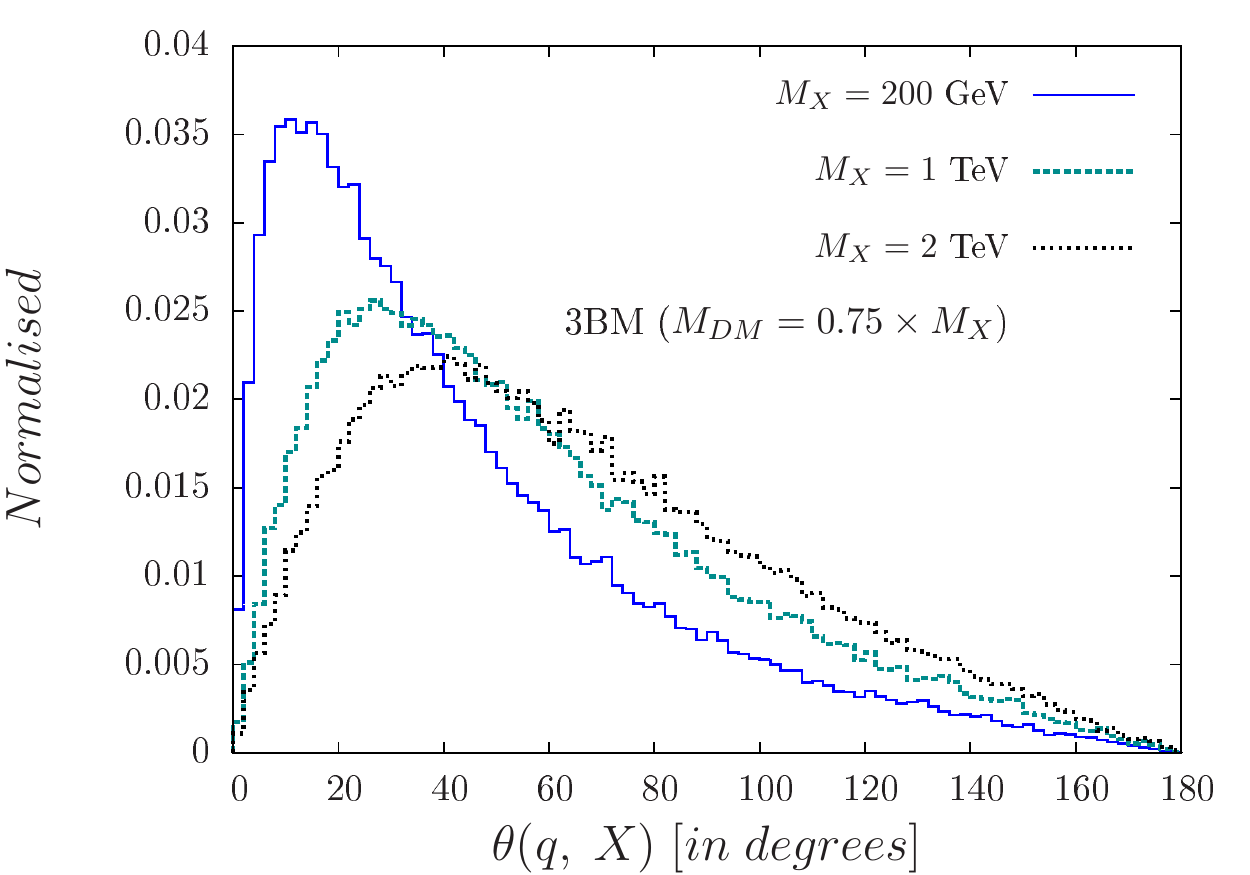}
} 
\subfigure[]
{
 \includegraphics[height=4cm,width=8.0cm]{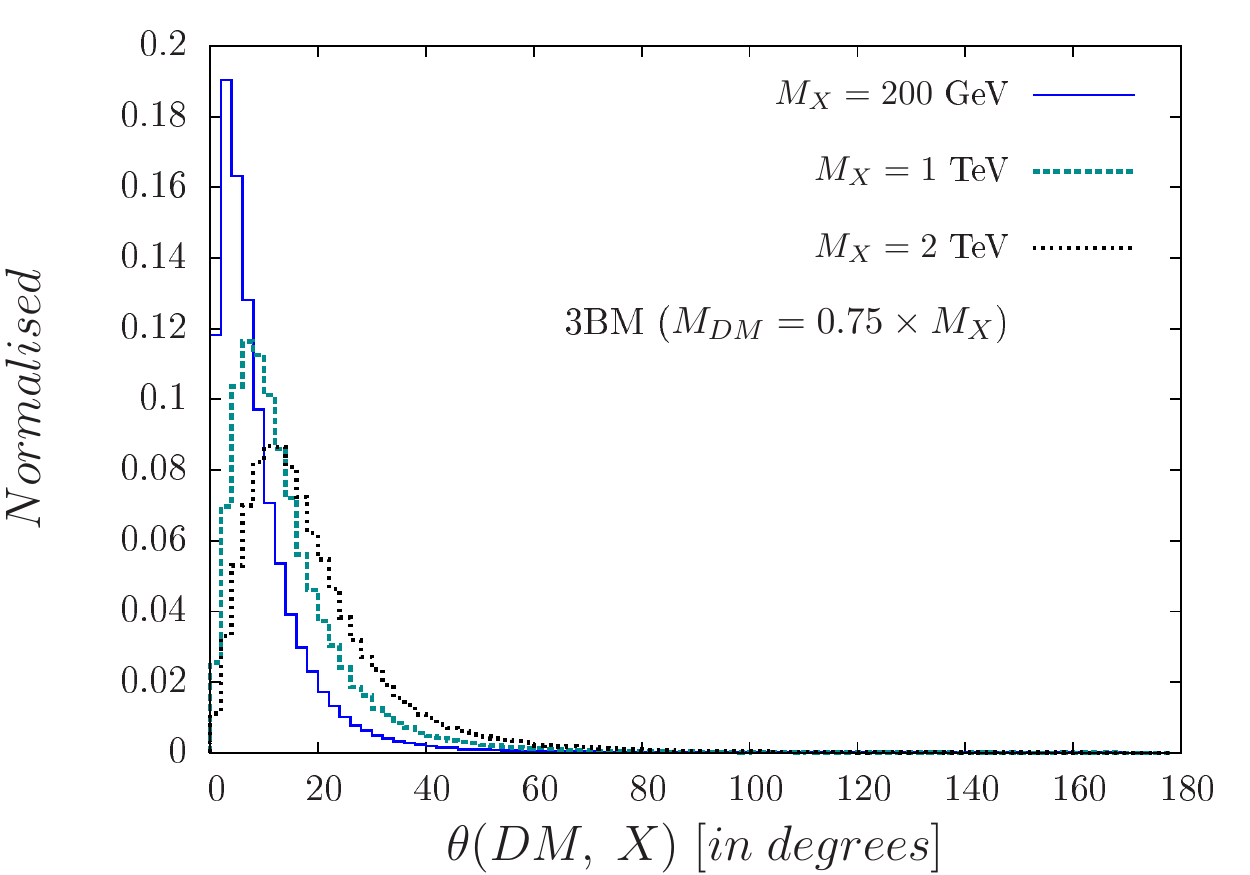}\\
} 
\subfigure[]
{
 \includegraphics[height=4cm,width=8.0cm]{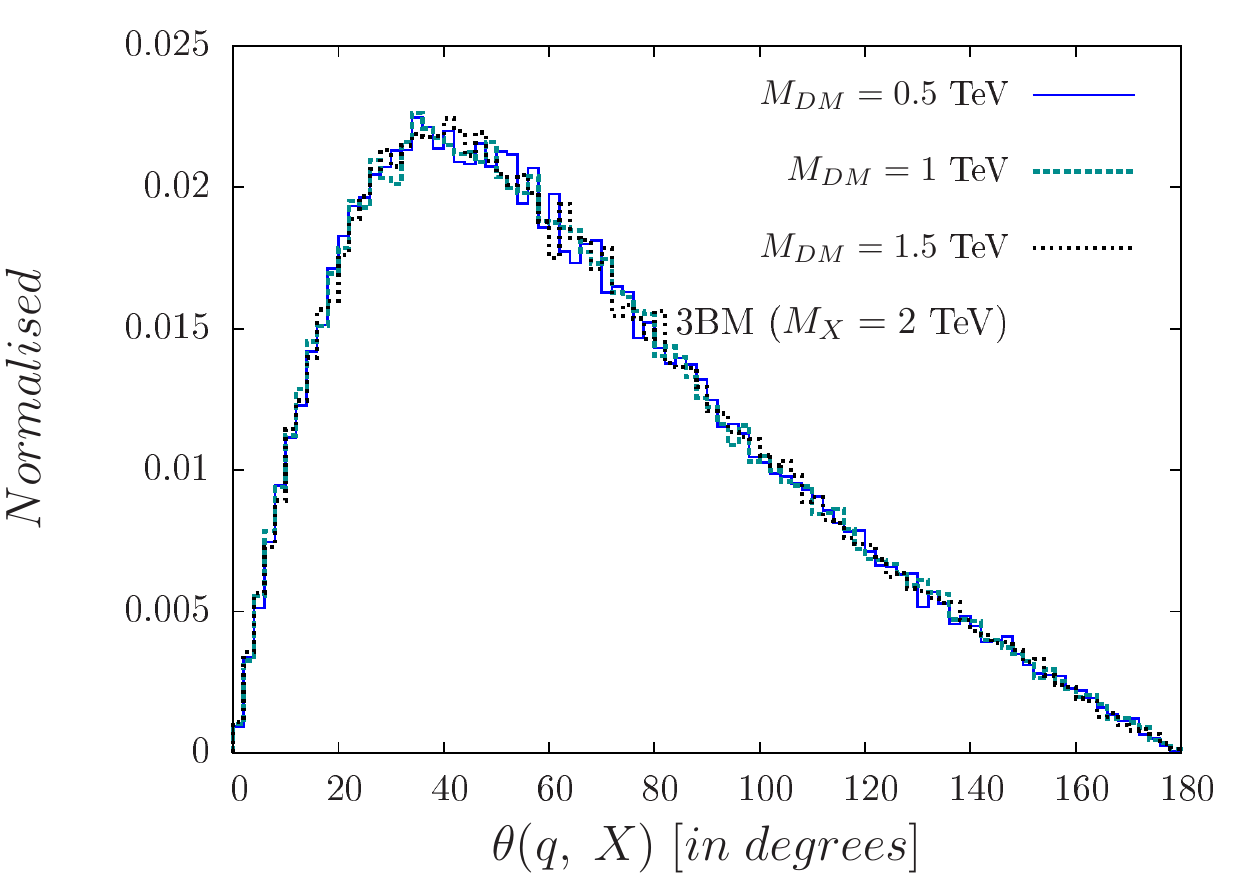}
} 
\subfigure[]
{
 \includegraphics[height=4cm,width=8.0cm]{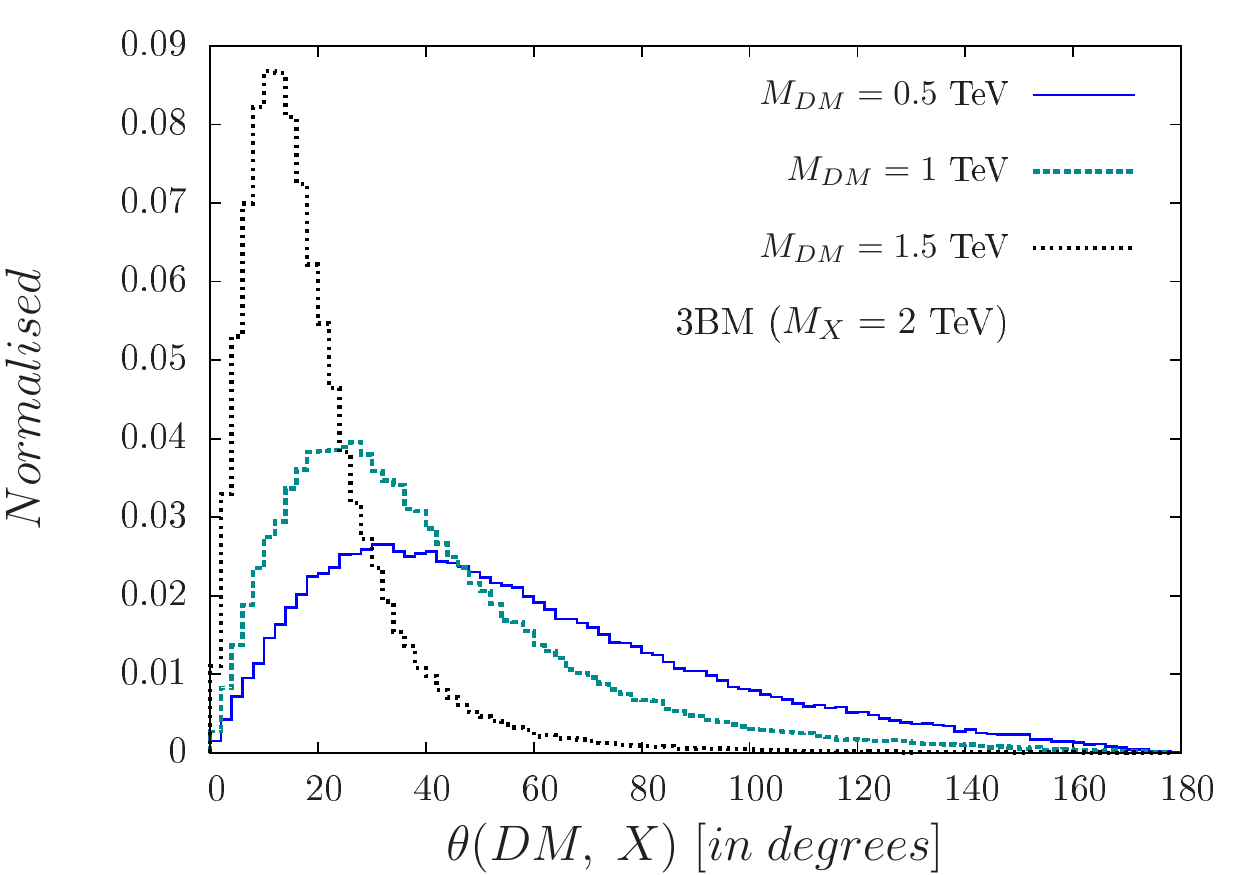}\\
} 
\caption{Angle $\theta$ between the direction of $X$ and the massless daughter (one of the quarks, $q$) or the massive daughter (DM) for the four scenarios (2BM0, 3BM0, 2BM and 3BM) for different values of $M_X$ and $M_{\text DM}$, as shown in the figure.}
\label{fig:theta_2_TeV}
\end{figure}

The above examples are illustrative and have been used to simulate our Monte Carlo samples. However, it is important to remember that the actual results that we will discuss in the following sections will be mostly model-independent. All these possibilities give rise to final states with multiple jets and, in the case of {\it 2BM} and {\it 3BM}, these jets are accompanied with missing transverse energy ($\slashed{E}_T$). To weigh the robustness of our findings, the examples we have taken cover both $qq$ initiated ({\it 2BM0} and {\it 3BM0}) and $gg$ initiated processes. To see the effects of the full spin-correlation, we consider the case where $X$ is a fermion, in the {\it 3BM0} and {\it 3BM} scenarios. Moreover, in order to see whether there is any bias that is introduced by a particular choice of our prototype simulation on the dynamics of the model, we compare the results of a full simulation (in the case of {\it 3BM}, with and without including spin-correlations) with those assuming no dynamics in the production, that is by considering a unit value for the matrix element (${\cal M}$). This helps us in finding out whether or not the results are mostly kinematics driven. This is also the reason we consider three distinct values for the mass of $X$, $M_X$, and for the same $M_X$ we also test two values for the mass of $\textrm{DM}$, $M_{\textrm{DM}}$. In this first investigation, all simulations are performed at the parton level for the 14 TeV LHC using \texttt{PYTHIA 6}~\cite{Sjostrand:2006za}. 

The important feature that we want to portray is through the angular distributions of the decay products, in particular the {\em observable} massless quark, with respect to the direction of the long-lived mother particle, $X$. We commence by studying the specific 
processes that we have introduced earlier without considering the spin information of $X$ in the decay. 
We then investigate the model dependence and the effects of the spin. The latter will be shown to be negligible.

Figure~\ref{fig:theta_2_TeV} shows the angle the \textit{massless} (and massive DM-like) decay particles make with the direction of motion of $X$~\footnote{Because the samples have been generated for SUSY processes using \texttt{PYTHIA 6}, there is no spin information and hence for the full \textit{massless} scenario, the angular distribution for all the daughter particles are identical.}. \\

As expected, for light mother particles ($M_X =200$ GeV), the decay products are preferentially highly boosted, becoming slightly less so if there is a heavy DM particle among the decay products, see Figures~\ref{fig:theta_2_TeV}(a-f). As the mass of the parent particle increases, the fraction of massless daughters that are emitted opposite to the direction of motion of the parent particle, {\it i.e.} backwards, gets larger and larger. However, one can clearly observe that, even for lighter masses of the parent particle, the fraction of massless daughter particles with $\theta (q,X) > 90^{\circ}$, is not negligible. For the largest mass of the decaying particle considered in this work, $M_X=2$ TeV, the distribution in the angle of the massless quarks is practically independent of the presence of a massive (DM) particle among the decay products. To summarise at this point, the message is that, independent of the channels and the specific dynamics, there is a non-negligible fraction of {\em backward} massless particles. This fraction increases with the mass of the mother particle since this is associated with a smaller $\beta$. Although for lighter masses of the mother particle the backward fraction is small,  in terms of total events, this is compensated  by the larger $pp \to XX$ cross-section.

\begin{table}[!t]
\scriptsize
\centering	
\begin{tabular}{+l^l^l^l^l^l^l^l}
\hline 
Case  & $M_{X}$ & $M_\textrm{DM}$ & $\beta$ (mean, RMS) & $\theta > 22.5^{\circ}$ & $\theta > 45^{\circ}$ & $\theta > 90^{\circ}$ & $\theta > 135^{\circ}$ \\ 
  &  [TeV] & [TeV] &  &  &  & & \\
\hline
\hline
2BM0 & 0.2 & - & 0.75, 0.23 & 0.85                    & 0.62                  & 0.25                  & 0.05 \\
\rowstyle{\itshape}%
     &     &   & 0.87, 0.13 & 0.78                    & 0.46                  & 0.13                  & 0.03 \\
     & 0.5 & - & 0.66, 0.24 & 0.96                    & 0.78                  & 0.33                  & 0.07 \\
\rowstyle{\itshape}%
     &     &   & 0.81, 0.14 & 0.94                    & 0.65                  & 0.19                  & 0.04 \\
     & 1   & - & 0.58, 0.23 & 0.99                    & 0.90                  & 0.42                  & 0.09 \\
\rowstyle{\itshape}%
     &     &   & 0.72, 0.15 & 0.99                    & 0.83                  & 0.28                  & 0.06 \\
     & 2   & - & 0.46, 0.20 & 1.00                    & 0.98                  & 0.54                  & 0.13 \\
\rowstyle{\itshape}%
     &     &   & 0.60, 0.14 & 1.00                    & 0.97                  & 0.40                  & 0.08 \\
\hline 
2BM  & 0.2 & 0.05  & 0.67, 0.24 & 0.73                    & 0.47                  & 0.16                  & 0.04 \\ 
\rowstyle{\itshape}%
     &     &       & 0.74, 0.21 & 0.67                    & 0.40                  & 0.13                  & 0.03 \\ 
     & 0.2 & 0.15  & 0.67, 0.24 & 0.73                    & 0.46                  & 0.16                  & 0.04 \\ 
\rowstyle{\itshape}%
     &     &       & 0.74, 0.21 & 0.67                    & 0.40                  & 0.13                  & 0.03 \\ 
     & 0.5 & 0.125 & 0.60, 0.23 & 0.80                    & 0.54                  & 0.20                  & 0.05 \\
\rowstyle{\itshape}%
     &     &       & 0.66, 0.21 & 0.78                    & 0.50                  & 0.17                  & 0.04 \\
     & 0.5 & 0.375 & 0.60, 0.23 & 0.80                    & 0.54                  & 0.20                  & 0.04 \\
\rowstyle{\itshape}%
     &     &       & 0.66, 0.21 & 0.77                    & 0.50                  & 0.17                  & 0.04 \\
     & 1   & 0.25  & 0.52, 0.22 & 0.85                    & 0.61                  & 0.24                  & 0.05 \\
\rowstyle{\itshape}%
     &     &       & 0.57, 0.19 & 0.84                    & 0.58                  & 0.21                  & 0.05 \\
     & 1   & 0.75  & 0.53, 0.22 & 0.85                    & 0.61                  & 0.24                  & 0.05 \\
\rowstyle{\itshape}%
     &     &       & 0.57, 0.19 & 0.84                    & 0.58                  & 0.21                  & 0.05 \\
     & 2   & 0.50  & 0.42, 0.19 & 0.90                    & 0.68                  & 0.29                  & 0.07 \\
\rowstyle{\itshape}%
     &     &       & 0.46, 0.17 & 0.89                    & 0.66                  & 0.27                  & 0.06 \\
     & 2   & 1.50  & 0.42, 0.19 & 0.90                    & 0.68                  & 0.29                  & 0.07 \\
\rowstyle{\itshape}%
     &     &       & 0.46, 0.17 & 0.89                    & 0.66                  & 0.27                  & 0.06 \\
\hline 
3BM0 & 0.2 & - & 0.76, 0.23 & 0.89                    & 0.69                  & 0.32                  & 0.07 \\
\rowstyle{\itshape}%
     &     &   & 0.94, 0.09 & 0.65                    & 0.34                  & 0.09                  & 0.02 \\
     & 0.5 & - & 0.67, 0.23 & 0.98                    & 0.84                  & 0.43                  & 0.10 \\ 
\rowstyle{\itshape}%
     &     &   & 0.86, 0.13 & 0.92                    & 0.61                  & 0.20                  & 0.04 \\ 
     & 1   & - & 0.58, 0.23 & 0.99                    & 0.94                  & 0.54                  & 0.14 \\
\rowstyle{\itshape}%
     &     &   & 0.76, 0.15 & 0.99                    & 0.84                  & 0.33                  & 0.07 \\
     & 2   & - & 0.46, 0.20 & 1.00                    & 0.99                  & 0.68                  & 0.18 \\
\rowstyle{\itshape}%
     &     &   & 0.62, 0.15 & 1.00                    & 0.98                  & 0.52                  & 0.12 \\
\hline      
3BM  & 0.2 & 0.05  & 0.67, 0.24 & 0.91                    & 0.70                  & 0.31                  & 0.07 \\
\rowstyle{\itshape}%
     &     &       & 0.76, 0.19 & 0.86                    & 0.60                  & 0.22                  & 0.05 \\
     & 0.2 & 0.15  & 0.67, 0.24 & 0.89                    & 0.67                  & 0.30                  & 0.07 \\
\rowstyle{\itshape}%
     &     &       & 0.77, 0.19 & 0.84                    & 0.58                  & 0.21                  & 0.05 \\
     & 0.5 & 0.125 & 0.60, 0.23 & 0.96                    & 0.79                  & 0.37                  & 0.09 \\  
\rowstyle{\itshape}%
     &     &       & 0.69, 0.19 & 0.94                    & 0.73                  & 0.29                  & 0.06 \\  
     & 0.5 & 0.375 & 0.60, 0.23 & 0.94                    & 0.76                  & 0.36                  & 0.09 \\  
\rowstyle{\itshape}%
     &     &       & 0.69, 0.19 & 0.92                    & 0.70                  & 0.28                  & 0.06 \\  
     & 1   & 0.25  & 0.53, 0.22 & 0.98                    & 0.86                  & 0.43                  & 0.11 \\
\rowstyle{\itshape}%
     &     &       & 0.61, 0.18 & 0.97                    & 0.82                  & 0.36                  & 0.08 \\
     & 1   & 0.75  & 0.52, 0.22 & 0.97                    & 0.83                  & 0.42                  & 0.10 \\     
\rowstyle{\itshape}%
     &     &       & 0.61, 0.18 & 0.96                    & 0.79                  & 0.35                  & 0.08 \\     
     & 2   & 0.50  & 0.42, 0.19 & 0.99                    & 0.93                  & 0.52                  & 0.13 \\ 
\rowstyle{\itshape}%
     &     &       & 0.50, 0.16 & 0.99                    & 0.90                  & 0.46                  & 0.11 \\ 
     & 2   & 1.50  & 0.42, 0.19 & 0.99                    & 0.90                  & 0.49                  & 0.13 \\      
\rowstyle{\itshape}%
     &     &       & 0.50, 0.16 & 0.98                    & 0.87                  & 0.44                  & 0.11 \\      
\hline 
\end{tabular} 
\caption{Mean value and dispersion (rms) of the velocity of the mother particle ($X$) and fraction of events with angle $\theta$ made by at least one of the lightest daughter particles with the direction of $X$, for the four scenarios. For each $M_X$ (and $M_{\text DM}$), we also give the ${\cal M}=1$ (kinematics only) case (first row). The row just below (in \textit{italics}) is for the the model-dependent scenarios.}
\label{tab:theta_unit}
\end{table}

Table~\ref{tab:theta_unit} makes the correlation with the boost of the mother particle, $X$, more apparent by showing the mean value of its velocity, $\beta$, and the associated fraction of massless decay particles (the quarks here which will be tracking the LLP) that are emitted at different angles relative to $X$. Four sectors in angular separation are defined. The most interesting are the ones where the daughter particle is emitted {\em backward}, {\it i.e.}, with $\theta > 90^{\circ}$. The table also shows the corresponding values for a matrix element ${\cal M}=1$ scenario, that is, a model driven solely by kinematics. First of all, $\beta$ is independent of the decay channel and depends only on the production process. For ${\cal M}=1$, we expect that the mean $\beta$ is the same between {\it 2BM0} and {\it 3BM0}, as it is the same between {\it 2BM} and {\it 3BM}, for the same mass (independently of $M_{\text DM}$). The small difference (even smaller for larger $M_X$), between {\it 2BM0} and {\it 3BM0} on the one hand and {\it 2BM} and {\it 3BM} on the other hand, reflects the $qq$ {\it versus} $gg$ production. As expected, the $s$-channel $qq$ production leads to slightly larger values of the mean $\beta$. The reason we gear the discussion around the mean $\beta$ is because as $\beta$ decreases, the fraction of {\it backward} events increases~\footnote{Since we are considering the particle motion in the transverse direction, the velocity in the transverse direction, $\beta_T$ is a more pertinent quantity. We find  that the mean and rms values of $\beta_T$ do not vary much with the LLP mass. However, because these distributions are asymmetric, we find that the fourth moment (kurtosis) parameter plays a significant role in discriminating the $\beta_T$ distributions and increases with the mass.}. Of course $\beta$ decreases as the mass of the mother particle increases and independently of the model, the fraction of {\it backward} massless quarks increases. We observe that, for the same mass, there is some model dependence in the value of the mean $\beta$, and for all models, $\beta$ increases as compared to the pure kinematics case. The largest difference is seen in the case of $3BM0$. But, in all cases the difference gets smaller as the mass ($M_X$) increases and with it the fraction of {\em backward} moving quarks also increases. In all four cases, it suffices to look at the mean $\beta$ to guess the angular fraction at, for example, $\theta > 90^{\circ}$. For instance $\beta \sim 0.75$ occurs for ${\cal M}=1$ and $M_X=200$ GeV as well as for $M_X=1$ TeV in the dynamical {\it 2BM0} case, moreover both display similar backward fractions. Similarly for the {\it 2BM} case ($\beta \sim 0.6$), the {\it 3BM0} case ($\beta \sim 0.76$) and the {\it 3BM} case ($\beta \sim 0.6$), the values of $\beta$ correspond to different mass scenarios, yet they lead to similar angular fractions. In summary, independently of the channel and the model, we find fractions of backward particles of at least $10\%$ (for small $M_X$) to values as high as $68\%$ (for larger masses).

\begin{figure}[tbhp]
\subfigure[]
{
 \includegraphics[height=5cm,width=8.0cm]{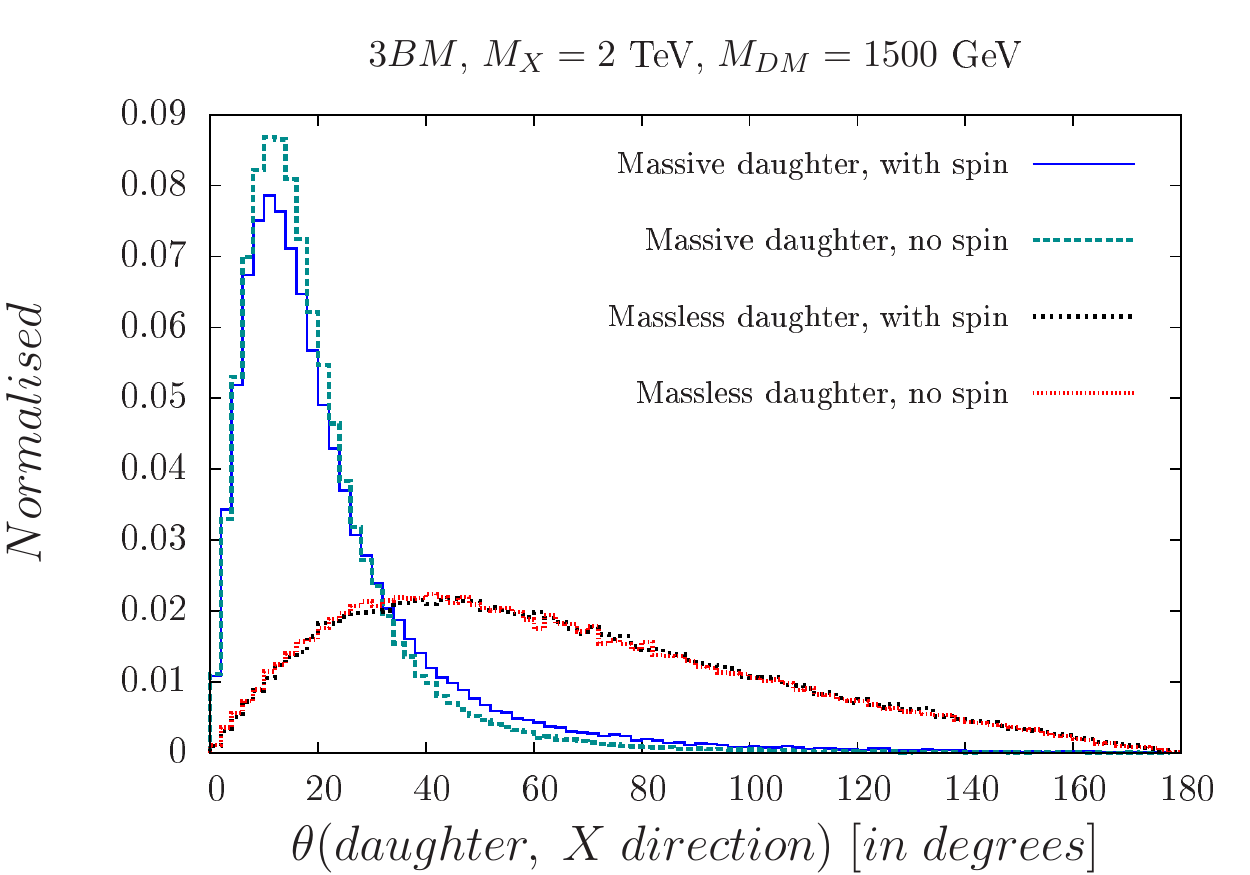}
}
\subfigure[]
{
\includegraphics[height=5cm,width=8.0cm]{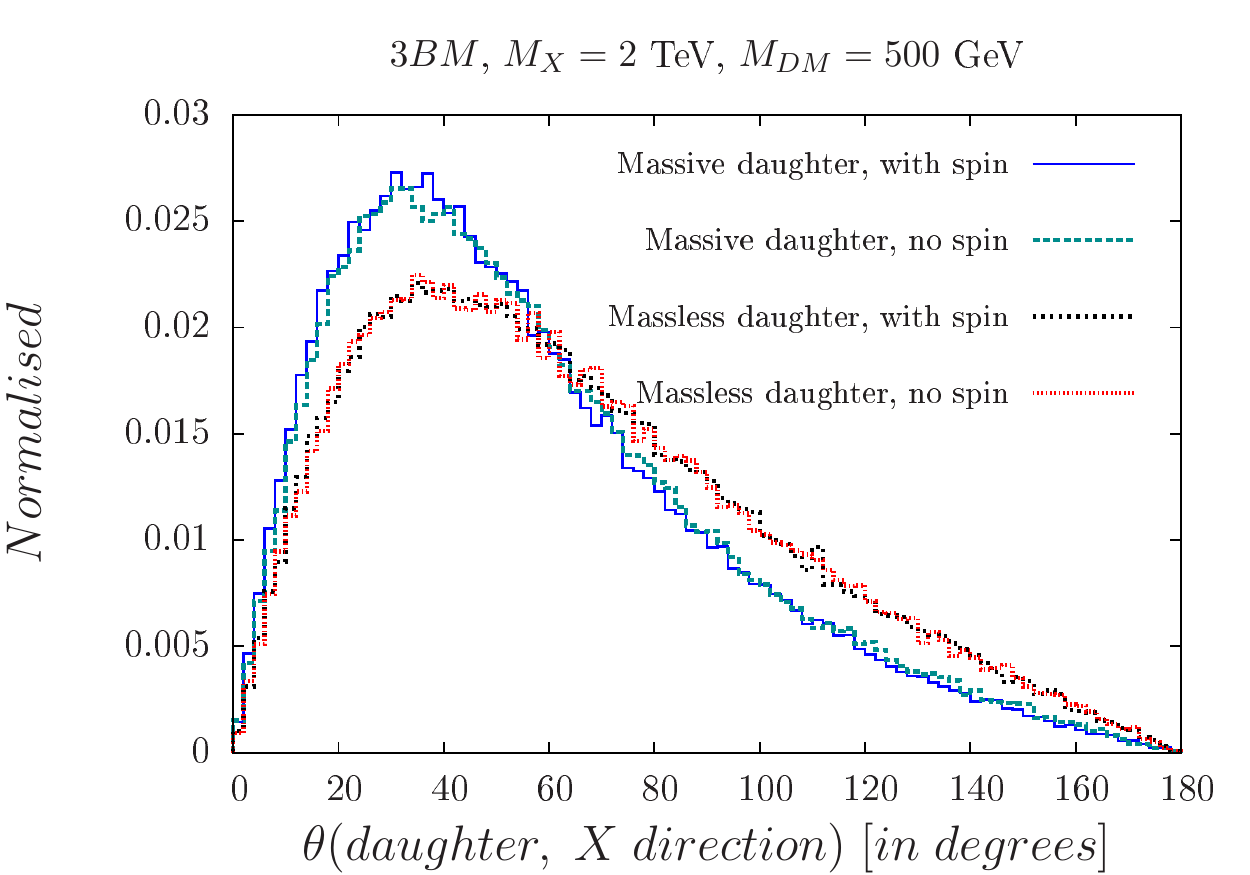}\\
} 
\caption{Angle $\theta$ made by a daughter particle with the direction of $X$ for (a) three body decays with one massive daughter with $M_X=2$ TeV and $M_{DM}=1.5$ TeV, (b) same with $M_{DM}=0.5$ TeV. Here we compare the simulation with full spin correlations and with the spin averaged approximation.}
 \label{fig:theta_2_TeV_spin}
\end{figure}

If {\em backwardness} is mostly driven by the velocity of the mother particle which in turn, is essentially driven by the kinematics of the initial state, we expect the spin of the mother particle to play a negligible part. To quantitatively check this, we consider three-body decays ($3BM$) and compare the approximation with the spin-averaged cross section with the full simulation taking into account complete spin correlations between the production and decay (we simulate full spin correlations with \texttt{MG5\_aMC@NLO}~\cite{Alwall:2014hca}). To make the point, we only consider a single benchmark scenario with $M_{LLP}=2$ TeV with two values of $M_{DM}$, \textit{viz.}, 500 GeV and 1.5 TeV. Figure~\ref{fig:theta_2_TeV_spin} shows that the angular distributions of the massless daughters are extremely well reproduced by the spin-averaged approximation. For the distribution of the massive (invisible) daughter, the approximation shows a slight difference. Therefore, for the rest of this study we work with the spin-averaged scenarios.

To sum up this discussion, we wish to underline that the more sluggish the mother LLP, the more important it is for us to study particles moving in the backward direction with respect to the direction of $X$. The extreme scenario is where $\beta$ of the mother particle becomes zero. Such scenarios can come about in stopped $R$-hadrons which move inside the detector up to a certain distance and then come to a standstill. 

In our analysis, the visible decay products have been assumed to be quarks. We could have just as well considered the LLP decays into leptons. The general feature of {\em outside-in} objects will remain unchanged, although in the analysis the hadronisation step will be different.

\section{Implications for LLP searches}
\label{sec:sec3}

Although, such angular distributions are well-known, their implications for LLP searches have not been thoroughly investigated until now. An LLP, upon production, moves a certain distance inside the detector and decays at a secondary vertex. We have learned that, especially for quite massive LLPs, there is a non-negligible proportion of the decay products that will not carry on in the original direction of the mother particle. For starters, the decay particles will not point in the direction of the interaction point. Depending on the angular separation of the \textit{visible} daughter with respect to the direction of the LLP, the decay products may reveal an {\em outside-in} activity in different parts of the detector. For example, non-prompt jets emanating from a secondary vertex could pass through multiple calorimeter towers yielding elliptical energy deposition in the $\eta-\phi$ plane of the HCAL. This is in contrast to {\em normal} jets born at the primary vertex which are usually contained within a single tower of the HCAL and yield a circular energy deposition. Similar energy distributions in the ECAL are expected from prompt and non-prompt photons~\cite{Chatrchyan:2012jwg}. We will not dwell further on these distorted objects (\textit{DOs}) because we would like to study the interesting case of the backward moving objects, \textit{BMOs}.

If the separation angle between the direction of the daughter and that of the mother is sufficiently large, this means that the daughter particle is moving in the backward direction. It can therefore  even cross inward layers of the detector (which a stable mother would not have done!). For example, if an LLP decays in the ECAL, the \textit{BMOs} will tend to move towards the tracker. As discussed in the introduction, this statement can be generalised in the context of decay products of an LLP moving from any outward detector segment to an inner one. Such unusual signatures are indeed striking and suffer from very low backgrounds. For sure SM particles produced or initiated by $pp$ collisions do not contribute to such signatures. We will address the issue of potential backgrounds to the \textit{BMOs} from LLP decays, which consist  essentially of cosmic rays, in the next section. To the best of our knowledge, dedicated searches for such \textit{BMOs} are yet to be performed at the LHC.

We attempt two exploratory analyses with \textit{BMOs} based on a {\em simplified geometrical analysis}. A detailed simulation leading to more realistic significances would require us to know the geometry and response of the different components of the detector, which is outside the scope of this work. In the present paper, we look at two regions. In the first example, we consider the HCAL-tracker region and in the second example we are interested in the muon chamber as a collector of otherwise lost signals for LLP decaying outside the detector. The results we will show pertain to a single LLP. Since daughter particles moving in the backward direction can occur from either of the pair produced LLPs, the actual statistics (and the significance) could therefore be larger. We approximately follow the dimensions of the CMS detector~\cite{Ball:2007zza} to quantify our analyses. The results can be generalised to the ATLAS detector. We exploit the 2BM/2BM0 and 3BM/3BM0 signatures defined in the previous section with hadronisation performed within the \texttt{PYTHIA 6} framework. We compute the ratio of the energy carried by the visible (hadronised) \textit{BMOs} that inwardly traverse the volume of interest,
 $E_{\textrm{in}}$, to the initial energy carried by the LLP, $E_{\textrm{LLP}}$. $E_{\textrm{in}}/E_{\textrm{LLP}}$ will be the characterising variable in our analysis. We expect this variable, with the consideration of the size of the particular layer of the detector (tracker, muon chamber), to still reflect the proportion of backward moving, {\em outside-in}, objects   as given in Table~\ref{tab:theta_unit}.

In both the tracker and the muon chamber application, we will consider the case of an LLP decaying in flight as well as the case of a stopped $R$-hadron. In both cases we take the same  mass and the same decay products. It is, of course, also assumed that both   decay in the  same region of interest. Naturally, the life-time of the LLP is assumed to be appropriate so as to yield a significant number of events within the region of interest. We keep this discussion model independent and do not make the exact lifetime explicit, nor the total cross section, as the following results will be fairly independent of these assumptions. For instance, the couplings of the underlying model can be easily tuned to get the desired lifetime. In this analysis, we are not attempting a precise modelling of the $R$-hadron's hadronisation as they move through the detector, yet we should  reproduce the main features of the $R$-hadrons.
In a sense, we are considering a toy skeleton of stopped $R$-hadrons to which we are looking at after they have come to a rest. We boost back all the daughter particles of that particular LLP to the stopped $R$-hadron's rest frame and compute the fraction of energy carried by them in the backward direction and inside that chosen layer of the detector (tracker in the first case and the muon chamber in the second example). 

At this stage, before we present the results of what we called our {\em simplified geometrical analysis}, we would like to issue an important warning. The analysis does not address crucial points about the reconstruction and the measurement of some key quantities. For instance, the specifics of the particular portion of the detector is not addressed. In this section, we will not discuss the response of a particular element of the detector to the \textit{BMO} and the identification of this \textit{BMO}. For example an identification and/or discrimination based on the possibility of timing or the shape of the showers, involves different issues depending on the location of the specific layer of the detector. In this section, we do not address how the energies we have introduced can be measured experimentally. For instance, reconstructing the secondary vertex in events with large impact parameters, is less trivial. As for $E_{\textrm{in}}$, the question of trigger may prove important. In this analysis, we have used $E_{\textrm{LLP}}$  mainly to express our results in terms of a normalised quantity,  $E_{\textrm{in}}/E_{\textrm{LLP}}$ rather than $E_{\textrm{in}}$. In the case where the decay involves invisible particles, we could substitute $E_{\textrm{LLP}}$ with the total transverse energy, $E_T$. It rests that the important discriminating observable that must be measured experimentally is $E_{\textrm{in}}$. In some cases, $E_{\textrm{LLP}}$, like for a neutral LLP, may not be measured but $E_{\textrm{in}}$ can bring invaluable information. We will come back to these very important points in section~\ref{sec:sec4}. Let us now return to our simplified geometrical analysis. 

\subsection{Reversing into the tracker}
\label{subsec:tracker}

We consider the tracker as an open cylinder having a length, $L_{\textrm{tracker}}=600$ cm, along the $z$-direction and a radius, $R_{\textrm{tracker}}=100$ cm. The last layer of the HCAL is considered to be at a transverse distance of 300 cm from the $z$-axis. For simplicity, our considerations pertain to the barrel only. The results can be  extended by including the end-caps. We compute the fraction of energy carried by  particles moving  from somewhere between the  outer edge of the HCAL as they make their way into the tracker volume. To do so, we employ a trivial geometry concerning a ray crossing a finite open cylinder. If the LLP decays between 100 cm and 300 cm in the transverse direction between the HCAL and the tracker we compute the fraction, $E_{\textrm{in}}/E_{\textrm{LLP}}$.
 
\begin{figure}[!h]
 \centering
 \includegraphics[height=4cm,width=8cm]{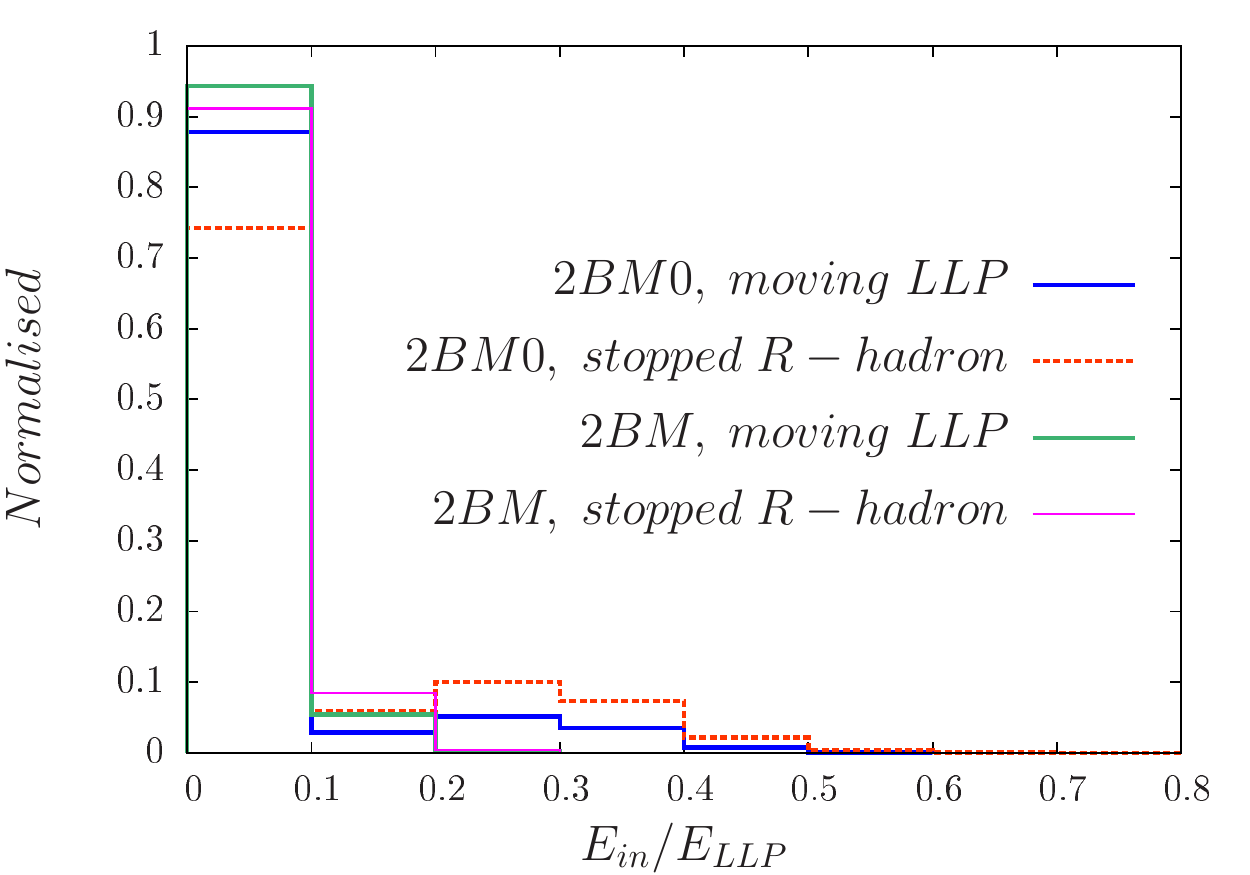}~\includegraphics[height=4cm,width=8cm]{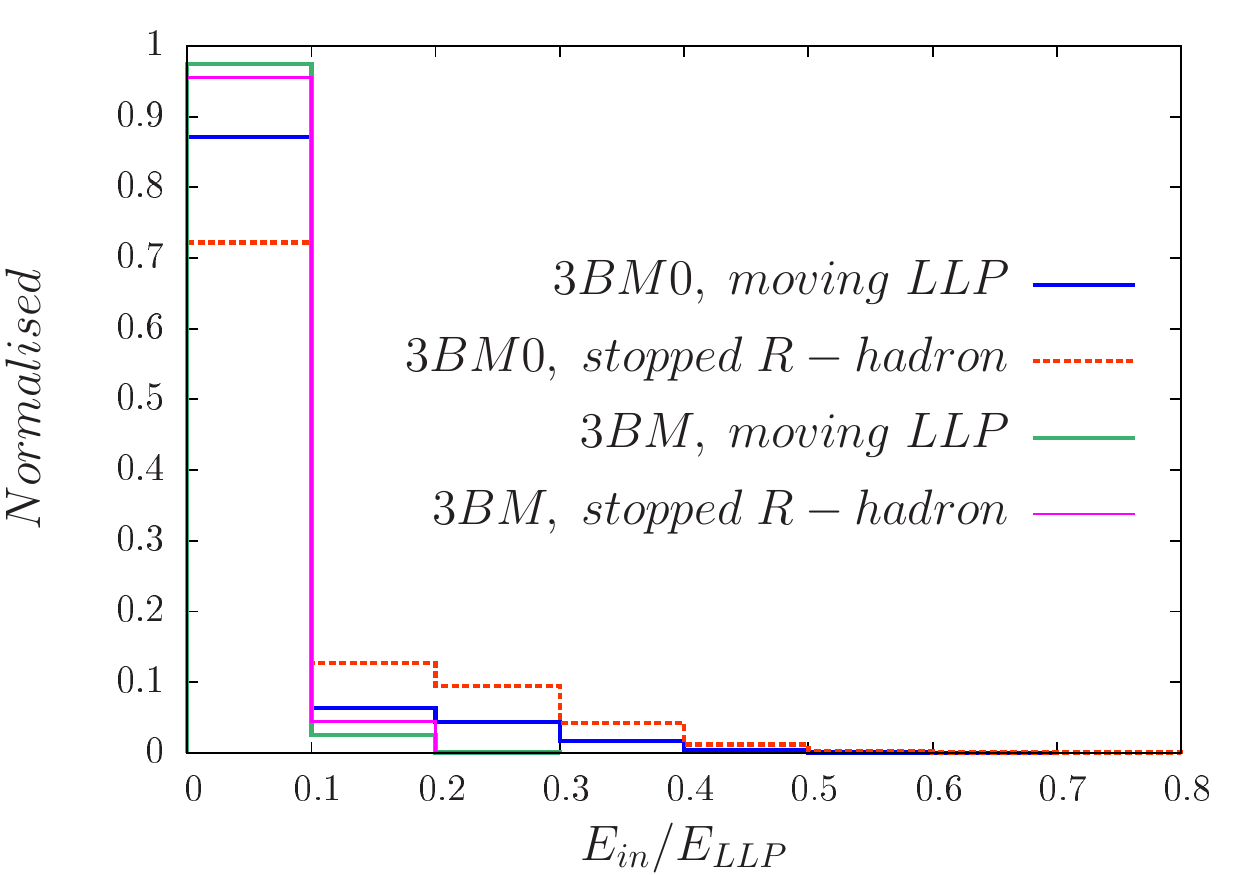}
 \caption{Normalised distribution of $E_{\textrm{in}}/E_{\textrm{LLP}}$, the energy fraction of visible daughter particles to the mother LLP shown for $M_{LLP}=2$ TeV and  $M_{\textrm{DM}}=0.75 \times M_{\textrm{LLP}}=1.5$ TeV. For the definition of the 2BM/3BM decays, see the text. In the first bin ($E_{\textrm{in}}/E_{\textrm{LLP}}< 0.1$) $E_{\textrm{in}}=0$. It should be interpreted as the case where no \textit{BMO} has registered.}
 \label{fig:Efac}
\end{figure}

Figure~\ref{fig:Efac}, shows the (normalised) distribution $E_{\textrm{in}}/E_{\textrm{LLP}}$ for a $2$ TeV LLP. While a large proportion of the LLP decay products do not make it into the tracker (these are represented by the first $E_{\textrm{in}}=0$ bin) independently of the decay channel, a substantial proportion does register inside the tracker as a signal for \textit{BMOs}. This proportion is larger for the stopped $R$-hadron case. These observations are in line with those we made in section~\ref{sec:sec2} based on the velocity of the LLP. This distinction is striking for the case when all the daughters are \textit{massless} (two-body 2BM0 and three-body 3BM0). For such scenarios, the fractions of energy coming back inside the tracker in the case of massless two-body decay is 25.9\% for the stopped $R$-hadrons and slightly less than half that number, 12.2\% for the moving LLP. In the case of three-body decays, these figures are slightly higher, respectively 34.2\% and 14.2\%. When one of the daughters is a massive invisible particle, the situation changes drastically, especially in the case of the $R$-hadron. The heavy daughter moves forward mostly in the direction of the mother LLP (as shown in figure~\ref{fig:theta_2_TeV}). The energy fractions traversing back into the tracker become, for the $R$-hadron,  8.2\% in the 2BM case and 4.6\% in the 3BM case. For the  corresponding moving LLP, we obtain  5.1\% (2.5\%) for the two-body (three-body) decay of the LLP. Upon varying the mass of the heavy invisible daughter particle, we find that the fraction $E_{\textrm{in}}/E_{\textrm{LLP}}$ changes appreciably. As an example, for $M_{\textrm{LLP}}=2$ TeV, for the 2BM decay mode, this fraction decreases approximately linearly from 8.5\% to 5.1\% upon changing $M_{\textrm{DM}}/M_{\textrm{LLP}}$ from 0\% to 75\%. We should keep in mind that these unconventional signatures have almost no SM background. Therefore, even though the $E_{\textrm{in}}$ fractions are smaller in scenarios where decay products of LLPs include massive invisible particles than in scenarios where the decay products consist exclusively of massless visible particles, the results we obtain are encouraging.

\subsection{Back into the muon chamber}

\begin{figure}[!h]
 \centering
 \includegraphics[height=4cm,width=8cm]{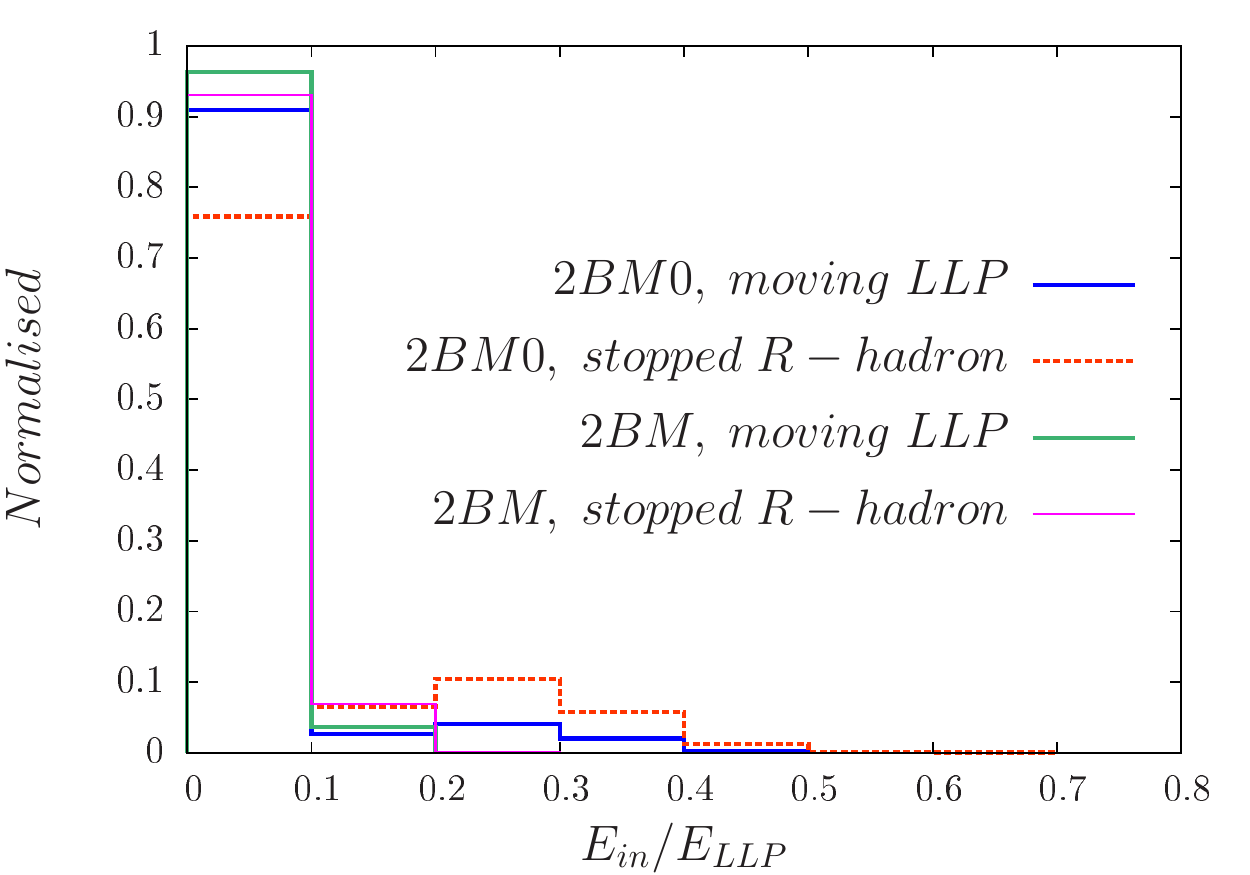}~\includegraphics[height=4cm,width=8cm]{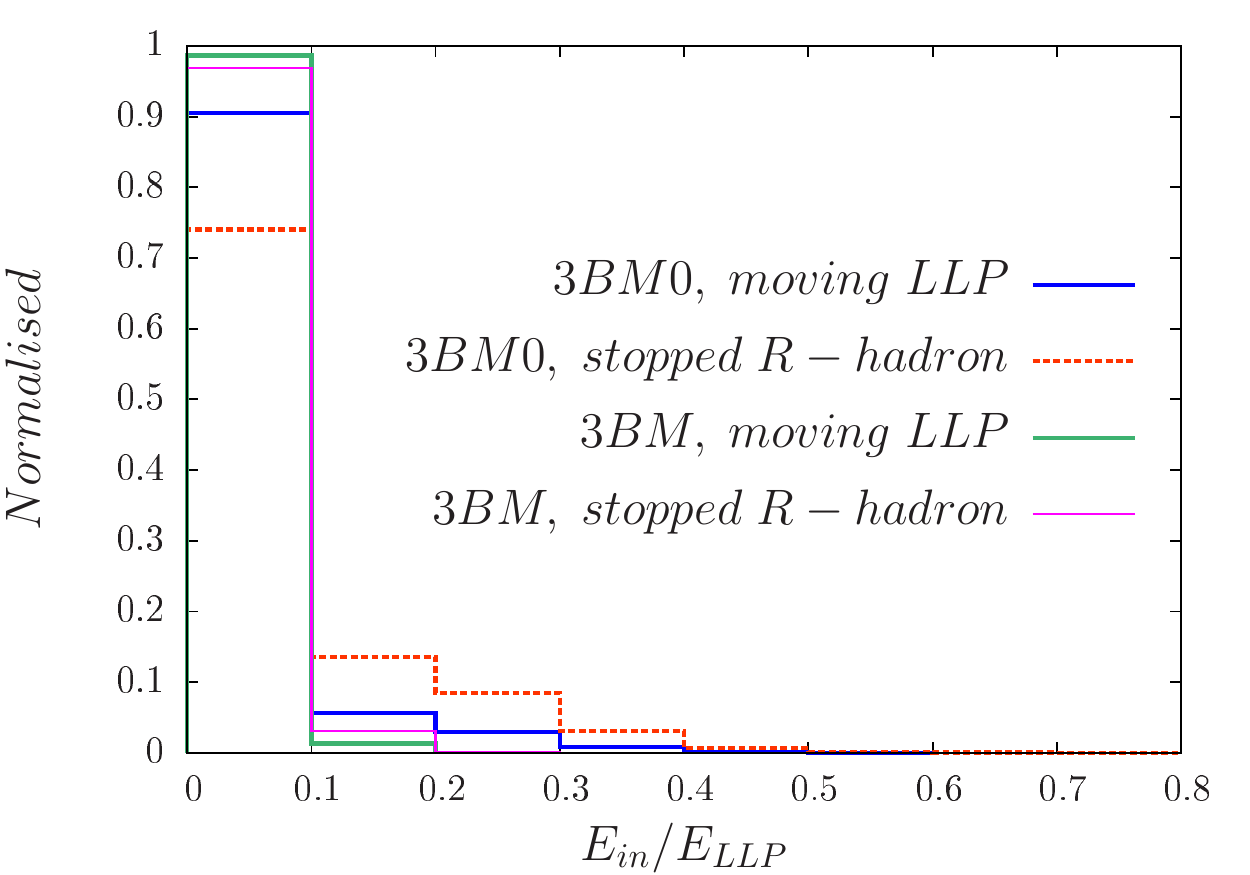}
 \caption{As in Fig.~\ref{fig:Efac} but for the case of the muon chamber with dimensions as specified in the text.}
 \label{fig:Efac-mu}
\end{figure}

To quantify a more concrete advantage of this framework, we consider particles that decay just outside the muon chamber. The only way of detecting such particles (inside the same detector) is if the daughter particles move inward towards the muon chamber. Here we again refer to the CMS geometry~\cite{cms-geometry}. We consider the muon chamber as a finite open cylinder of radius, $R_{\textrm{muon-chamber}}=750$ cm and a length of $L_{\textrm{muon-chamber}}=1300$ cm along the $z$-direction. The CMS experimental cavern is around 26.5 m in diameter and the diameter of CMS is around 15 m. Hence, there is a volume between the CMS detector and the cavern which may not all be empty. We consider the LLP to decay outside the muon chamber, somewhere between 750 cm and 1500 cm. Finally, we compute the same fraction, \textit{viz.}, $E_{\textrm{in}}/E_{\textrm{LLP}}$ for the two-body and three-body decay scenarios. In figure~\ref{fig:Efac-mu}, we show these ratios for the cases with $M_{LLP}=2$ TeV and $M_{\textrm{DM}}=1.5$ TeV. \\

The fractions of energy coming back inside the muon chamber is similar to the energy fractions we calculated for the tracker, especially in the case where all decay particles are visible (2BM0/3BM0). In the case of the two-body decays, the fraction is as much as 24\% for the stopped $R$-hadrons but it is less than half that number, 9\%, for the moving LLP. In the case of three-body decays, these figures are slightly higher, respectively 26\% and 10\%. When one of the daughters is a massive invisible particle, there is an important deterioration, worse than what we observed in the case of the tracker, especially in the case of three-body decays. For the $R$-hadron, the percentages drop to 7\% for the 2-body and only 3\% for the 3-body decay scenario. For the moving LLP,  one has 4\% for the 2-body and only 1\% for the three-body. Even in this case, and depending on the statistics, let us not forget that this is one of the unique handles to resuscitate LLPs that decay outside the muon chamber.

\section{Experimental considerations and future upgrades}
\label{sec:sec4}

In this section we will discuss some of the important points that we left out in the previous section. After discussing about the background, we will turn to how the \textit{BMOs} can be tracked down (shower-shapes, timing, etc.). This very much depends on the specific slice of the detector (tracker, ECAL, HCAL, muon chamber). We will also briefly review how one can improve reconstruction and how future upgrades can help (whether it affects the secondary vertex reconstruction, new timers, etc.). 

\subsection{Backgrounds and background mitigation}

A \textit{BMO} signal is striking because it will not be recorded in the same pattern as that of the SM particles that originate from $pp$ collisions, making their way, transversally, from the beam-pipe to the outer-layers of the detector. There may be challenges coming from beam-induced noise, overlapping events (timing and/or shower shapes, see later, should help here) and instrumental noise. But by far, the most important background is the one not produced by the $pp$ machine, it is the one due to cosmic ray events~\cite{Gibson:2017igw, Rodenburg:2014urc, Aad:2016tzx}. This is particularly problematic when the signal is looked for in the muon chamber. One way to suppress such backgrounds is by tagging the backward moving LLP only in the lower half of the detector which will be almost free of any cosmic rays that move towards the beam-pipe. Exploiting events from the upper hemisphere can be a bit more challenging. To attempt giving any estimate for the cosmic muon background in the upper hemisphere requires knowledge of the event selection cuts. However, it is to be noted that the signature of hadrons in the muon detector is not a well studied subject. It needs to be checked, preferably by the experimental collaborations, using a proper full simulation. It is possible that a hadron in the muon chamber would be easily distinguishable from a cosmic muon in the muon chamber, because of the difference of signature. In that case the cosmic muon background will not be a big issue. If the  LLP decays in the tracker or the calorimeters, cosmic muons should not be a problem, and both the upper and lower hemispheres could be used. 

\subsection{Shower shapes for the ECAL}

Shower-shape for an {\em inside-out} jet is expected to be different from a backward-moving {\em outside-in} jet. There are widely used shower-shape variables for the ECAL, \textit{viz.}, $S_{major}$, $S_{minor}$, $\sigma_{i\eta i\eta}$ and $R_9$~\cite{CMS-PAS-EXO-12-035, CMS:ril, Khachatryan:2015qba}. The shape of the energy deposit in ECAL is characterised by the major and minor axes ($S_{major}$, $S_{minor}$), and of its projection on the internal ECAL surface. The variables $S_{major}$ and $S_{minor}$ are computed using the geometrical properties of the distribution of the energy deposit. The variable $\sigma_{i\eta i\eta}$ is the energy weighted standard deviation of single crystal $\eta$ within the $5 \times 5$ crystals centred at the crystal with maximum energy. The variable $R_9$ is the ratio of the energy deposited in the $3 \times 3$ crystal matrix surrounding the highest energy crystal to the total energy. For \textit{BMOs} decaying inside the ECAL, the aforementioned shower shape variables along with the ECAL timing information~\cite{CMS-PAS-EXO-12-035} can be utilised to distinguish such striking signatures and also to potentially reduce backgrounds.

\subsection{Shower shapes for the HCAL and calorimeter upgrades}

For signal signatures pertaining mostly to jets, let us discuss some possible shower-shape variables specific to the HCAL. This is particularly important when the LLP decays in one of the outer layers of the HCAL or even after crossing it and at least one of the decay products comes back inside the HCAL. If the HCAL has depth-segmentation then the energy of each depth can be read-out separately. If $E(D_i)$ denotes the energy deposited in the $i^{th}$ depth of a HCAL tower, then one can use $E(D_i)$ as inputs to train a boosted decision tree (BDT)~\cite{Roe:2004na}. The BDT output should be a powerful discriminator between backward-moving signal jets and forward-moving background jets.

After the phase II upgrade in 2024-2025, the CMS detector is expected to have a high-granularity calorimeter (HGCAL)~\cite{Martelli:2017qbe} in the forward direction, {\it i.e.}, towards the endcaps, which will have high-precision timing capabilities. The calorimeter design, with fine granularity in both lateral and longitudinal directions, is ideally suited to enhance such pattern recognition. Fine longitudinal granularity allows fine sampling of the longitudinal development of showers, providing good energy resolution, pattern recognition, and discrimination against pile-up. On the other hand, fine lateral granularity will help us to separate two close-by showers. After these improvements in the detector, the \textit{BMOs} in the forward part of the detector can be tagged more efficiently using the improved granularity and timing information.  
 
\subsection{Timing in the muon chamber and upgrades for the tracker}

If an LLP decays just outside the muon chamber, then the \textit{BMOs} are the only detectable objects in the signal. 
These \textit{BMOs} will reach the muon chambers two or more bunch-crossings after its production, and will give rise to signatures resembling that of late-muons. The CMS 
experiment has reported their trigger capabilities for such kind of exotic signatures in Refs.~\cite{latemuon1} and \cite{latemuon2}.
Moreover, in such cases, the timing information of the muon detectors (for example: resistive plate chambers in CMS) can be useful. Resistive plate chambers (RPC) are gaseous parallel-plate detectors that have good spatial resolution and excellent time resolution. The spatial resolution of RPC is of the order of 1 cm, and the time resolution is around 2-3 ns. So, it is capable of tagging the time of an ionising particle in a much shorter time than the 25 ns between two consecutive LHC bunch crossings. If $t_n$ is the timing of the hit in the $n^{th}$ layer of the muon detector~\footnote{Here, the innermost layer is assumed to be the first layer and $n$ increases as we move radially outwards.} for a reconstructed muon-track, then $t_n<t_{n+1}$ will be the signature of outward-moving background tracks and $t_n>t_{n+1}$ will be the sign of a \textit{BMO}. Something similar can not be done in the silicon tracker, because of its slow response time. However, the CMS collaboration is seriously considering the option of installing an additional timing layer~\cite{Josh} during the phase II upgrade of the detector in 2024-2026. This precise timing detector might sit just outside the tracker barrel support tube, in between the tracker and the ECAL barrel. This thin layer is expected to have a time resolution of 10-20 picosecond and it will provide timing for the individual tracks crossing it, while photon and neutral hadron timing will be provided by the upgraded calorimeters. The timing detector will be used to assign the timing for each reconstructed vertex and to measure the time of flight of the LLPs between the primary and secondary vertices. Thus, it would provide new, powerful information in searches for LLPs.

\subsection{Secondary vertex reconstruction, trackers and triggers}

\textit{BMOs}, that are heavily displaced with respect to the primary vertex, having large impact parameters, are likely to be missed by the currently used jet reconstruction algorithms, because such algorithms are based on the assumption that the jets are originating from the collision point. However, the jet reconstruction algorithm can be tuned to catch displaced jets. This option can be heavily resource-consuming and the experiments can utilise the ideas of data-scouting and parking~\cite{CMS-DP-2012-022}. Reconstructing \textit{BMOs}, with large impact parameters, inside the tracker, can be extremely challenging, but can be achieved by making modifications in track reconstruction algorithms, for example, by relaxing the requirement on the impact parameters of the track. One can use the concept of regional-tracking~\cite{Tosi}, \textit{i.e.}, the non-pointing tracks of only those regions of the tracker will be reconstructed where there is a corresponding calorimeter energy deposit. This concept is already used in track reconstruction in high-level trigger (HLT) in CMS. Reconstruction of tracks is a sophisticated, complex and time-consuming step. In order to make it faster during the data-taking at the HLT, some modifications have been done to the actual track reconstruction technique, that is used offline, which is not pressed by time. One of the modifications in order to save time, is to use regional track reconstruction, where tracking algorithm is run only in regions-of-interest defined by the direction of an already available physics object or calorimeter energy deposit. Even with modified track reconstruction techniques, it might be very difficult to distinguish between signal and background tracks. However, a recent study~\cite{Gershtein:2017tsv} has shown the capabilities of the high luminosity runs of the LHC (HL-LHC) in extracting more information from non-pointing tracks. A set of dedicated triggers might be needed to select such signal events within the LHC experiments. One possibility is to require multiple displaced jets with appropriate $p_T$ cuts. Otherwise, one can trigger on the sum of HCAL energy deposits. 

\section{Conclusions}
\label{sec:conclusions}

There has been in the last couple of years a rather intense activity in the search for long lived particles. The lack of any signal from the conventional searches of many BSM particles is one of the reasons behind this renewed interest. Because of their long life-time, the LLPs decay some distance away from the interaction point, at a secondary vertex, or even decay outside the detector. They may therefore easily be missed by standard searches. One should therefore leave no stone unturned and critically revisit any possible trace that they may leave on any sector of the detector, even if one can not trace back their production point. The main observation we make in this paper is that a, far from negligible, proportion of some of the visible decay products of the LLP will be moving {\em outside-in}, meaning that they will be moving from the location of the secondary vertex somewhere inside the detector towards the inner layers of the detector, in the direction of the beam pipe. These backward moving objects, \textit{BMOs}, will therefore have a most striking manifestation. It can even happen that the LLP may decay outside the detector but that some of its \textit{BMO} daughters will ``move back"  to deposit energy in the muon chamber. This crucial property of the \textit{BMOs} results from the fact that if the mother LLP is not too fast moving, these decay products will not be much boosted in the direction of flight. An extreme case is the one where the LLP decays at rest, and barring some spin effect, the decay products are distributed in all directions. If the LLP is sluggish at production, at the decay location some of its daughters will not carry on in the direction of the parent. In section~\ref{sec:sec2} we make this observation  quantitative when we study the angular separation that the visible decay product makes with respect to the direction of the parent LLP. We considered different scenarios and masses for the LLP as it is produced in pair at the LHC, either through a $q\bar q$ initiated or gluon-gluon initiated mechanism. We analysed, through the general models of LLP we introduced, the possible effect of the spin of the LLP, just to find out that spin effects are not important. We even make quantitative the expectation that the effect is mostly the result of kinematics, how slow the LLP is, and that the exact dynamics (the physics model dependence) is not crucial, by implementing a unit matrix element for the production. Although the model for the decay is inspired by some classes of LLP found in the literature, they cover essentially two classes. Either all decay products are visible or one of them is invisible (a possible Dark Matter candidate), in which case we investigate how heavy the latter is with respect to the parent LLP. As expected, the proportion of \textit{BMOs}, for example the fraction of visible daughters in a direction of more than $135^\circ$ from the original direction of the LLP is more substantial for larger masses of the LLP. This enhanced effect with higher masses should compensate the correspondingly smaller cross sections. It is therefore important to exploit this signature.   

For this simple analysis, we have only considered light jets as the visible objects. However, one can study other signatures involving leptons, photons or even boosted objects like top-jets, $W/Z/h$-jets. Performing a more realistic, let alone a full simulation, for this unusual signature would require detailed information on the different components of the various layers of the detector. Nonetheless, we have attempted to model the {\it geometry} of the tracker and the muon chamber (based on the dimension of the CMS sections of these layers) to quantify how the effect of a large angle separation translates into a measurable fraction of energy (with respect to the original energy of the LLP) that gets deposited respectively in the tracker from \textit{BMOs} emerging from as far as the HCAL and in the muon chamber for \textit{BMOs} entering from outside the detector. As expected, the largest energy deposits are for stopped $R$-hadrons and the smallest in cases where the phase space left for the visible objects is reduced by the presence of a large mass taken by the invisible particle present in the decay. The results we obtained could most probably be optimised by combining them with the use of other variables, like for instance the use of transverse energies or even better the knowledge of a specificity of the particular layer. We discuss some of these issues, either based on what is already implemented in the current detectors or what could be implemented in the future, to help better track the \textit{BMOs}. In particular, we review how the shower shapes of the ECAL and the HCAL could be exploited and optimised, together with the timing techniques in the muon chamber and the improvements we could have in the tracker. Another aspect which needs more attention, since it is somehow a defining characteristic of the LLP, is the reconstruction of the secondary vertex in case of large impact parameters. Many of the improvements may be in place in the high luminosity option of the LHC, which could help increase the signal statistics of the LLP. One should however pay special attention to techniques of mitigating the underlying events and their influence on the improvement of the timing information to decipher the LLP in some layers of the detector. We have also argued that the main background, from cosmic rays, can be eliminated. In the worst case, we can restrict the analysis to the lower half of the detector.

All in all, the proposal we make in this paper looks very promising for the search of the LLP at the LHC, especially for the quite massive ones (above $500$ GeV). As we have discussed, this preliminary study calls for the investigation of a wide range of theoretical, phenomenological and experimental issues and optimisations so we can take full advantage of all the runs of the LHC. \\

{\it \bf Acknowledgments ---}
We thank Sunanda Banerjee, Nicolas Berger, Gustaaf H. Brooijmans, Shilpi Jain, Remi Lafaye, Maurizio Pierini,  and Giacomo Polesello for useful discussions. This work was supported in part by the French ANR project DMAstro-LHC (ANR-12-BS05-0006), by the {\it Investissements d'avenir} Labex ENIGMASS,  by the Research Executive Agency of the European Union under the Grant Agreement PITN-GA2012-316704 (HiggsTools),  by the  CNRS LIA-THEP (Theoretical High Energy Physics) and the INFRE-HEPNET (IndoFrench Network on High Energy Physics) of CEFIPRA/IFCPAR (Indo-French Centre for the Promotion of Advanced Research). The work of SM is supported by the German Federal Ministry of Education and Research BMBF. The work of BB is supported by the Department of Science and Technology, Government of India, under the Grant Agreement number IFA13-PH-75 (INSPIRE Faculty Award). The work of RMG is supported by the Department of Science and Technology, India under Grant No. SR/S2/JCB-64/2007. BB acknowledges the hospitality of LAPTh where the major parts of this work were carried out. BB, SB, GB and FB acknowledge the Les Houches workshop series ``Physics at TeV colliders'' 2017 where the work was finalised. The work of SB is also supported by a Durham Junior Research Fellowship COFUNDed between Durham University and the European Union under grant agreement number 609412
\bibliographystyle{JHEP}
\bibliography{biblio}

\end{document}